%% file: manuscript.tex
\documentclass{scrartcl}

\input{arxiv}
\usepackage{subcaption}

\usepackage{dblfloatfix} 
\allowdisplaybreaks 

\DeclareMathOperator{\Tr}{Tr}
\renewcommand{\i}{\mathrm{i}}
\newcommand{\rmi}{\mathrm{i}}
\newcommand{\GC}{\mathcal{G}_\mathrm{C}}
\newcommand{\GL}{\mathcal{G}_\mathrm{L}}
\newcommand{\GR}{\mathcal{G}_\mathrm{R}}

\newcommand{\HC}{\mathcal{H}_\mathrm{C}}

\newcommand{\SL}{\Sigma_\mathrm{L}}
\newcommand{\SR}{\Sigma_\mathrm{R}}
\newcommand{\GaL}{\Gamma_\mathrm{L}}
\newcommand{\GaR}{\Gamma_\mathrm{R}}
\newcommand{\condmat}{\hat{G}}

\newcommand{\bi}[1]{\boldsymbol{\mathit{#1}}}
\newcommand{\bvarphi}{\boldsymbol{\varphi}}

\onecolumn 

\title{Quantum transport in graphene nanoribbon networks: complexity reduction by a network decimation algorithm}

\addauthor{Tom Simon Rodemund}{1}
\addauthor{Fabian Teichert}{1}
\addauthor{Martina Hentschel}{1}
\addauthor{J\"org Schuster}{2,3}

\addaddress{Institute of Physics, Technische Universit\"at Chemnitz, D-09107 Chemnitz, Germany}
\addaddress{Center for Materials, Architectures and Integration of Nanomembranes, Technische Universit\"at Chemnitz, D-09107 Chemnitz, Germany}
\addaddress{Fraunhofer Institute for Electronic Nano Systems, D-09107 Chemnitz, Germany}

\email{tom.rodemund@physik.tu-chemnitz.de}

\abstract{
We study electronic quantum transport in graphene nanoribbon (GNR) networks on mesoscopic length scales.
We focus on zigzag GNRs and investigate the conductance properties of statistical networks. 
To this end we use a density-functional-based tight-binding model to determine the electronic structure and quantum transport theory to calculate electronic transport properties.
We then introduce a new efficient network decimation algorithm that reduces the complexity in generic three-diemnsional GNR networks.
We compare our results to semi-classical calculations based on the nodal analysis approach and discuss the dependence of the conductance on network density and network size.
We show that a nodal analysis model cannot reproduce the quantum transport results nor their dependence on model parameters well. Thus, solving the quantum network by our efficient approach is mandatory for accurate modelling the electron transport through GNR networks.
}

\addkeyword{network decimation algorithm}
\addkeyword{graphene nanoribbon (GNR)}
\addkeyword{network}
\addkeyword{density-functional-based tight binding (DFTB)}
\addkeyword{quantum transport}
\addkeyword{recursive Green's function formalism (RGF)}
\addkeyword{decimation scheme}
\addkeyword{nodal analysis}
\addkeyword{mesoscopic transport}
\addkeyword{semiclassical transport}

\journal[]{New Journal of Physics 25 (2023), 013001}{https://iopscience.iop.org/article/10.1088/1367-2630/acaef0} 
\doi{10.1088/1367-2630/acaef0} 
\arxiv{2212.07238}{cond-mat.mes-hall} 

\begin{document}

\maketitle

\section{Introduction}

The progressing miniaturisation sets increasing demands on microelectronic devices like interconnects, transistors, and sensors.
They must be mechanically stable, i.e. material diffusion due to electromigration must be omitted~\cite{Nanotechnology.20.075706, IEEEElectronDevLett.33.1604}, and electronically stable~\cite{NanoLett.19.1460}, i.e. variations in its nanoscopic geometry must not negatively affect the device performance~\cite{NanoLett.6.2748}.
One approach to achieve this is the use of new low-dimensional, carbon-based materials like carbon nanotubes, graphene, or graphene nanoribbons (GNRs)~\cite{geim2009graphene, novoselov2012roadmap, aelm.202100358, IEEEElectronDevLett.33.1604, NanoLett.19.1460, NanoLett.6.2748}.
A challenging task is the calculation of material properties in the mesoscopic range with significant quantum effects where (semi-)classical descriptions break down, and quantum approaches come along with enormous computational cost.
Nevertheless, many quantum studies have been done using efficient recursive approaches~\cite{JPhysCSolidStatePhys.5.2845, CompPhysCommun.20.11, JPhysCSolidStatePhys.14.235, ZPhysBCondMat.59.385}, e.g. describing the influence of metal contacts~\cite{PhysRevB.77.125420, zienert2010transport, Nanotechnology.25.425203, PhysRevLett.119.207701} or imperfections~\cite{cresti2008charge, NanoLett.9.2725, JPhysCondMat.26.045303, PhysStatSolB.247.2962, NJPhys.16.123026, ComputMatSci.138.49, JPhysCommun.2.105012}.
For even larger systems like networks~\cite{ComputMatSci.161.364, rahman2012effects, lee20152d} or whole transistors~\cite{aelm.202100358} (semi)-classical approaches are used.

In this publication, we study GNR networks of different sizes and densities. We tackle the question how transferable quantum and (semi-)classical results are in the mesoscopic range where quantum effects may be relevant.
For this, we perform quantum transport (QT) calculations as well as nodal analysis (NA) in comparison. This work is organized as follows:
In section \sref{sec:QT} we first introduce the model system of the GNR network.
Second, we give a brief overview about the general basic QT equations.
Third, we present our new recursive network decimation scheme algorithm to efficiently calculate the key quantities.
We then extend the linear complexity scaling of the standard recursive algorithm for linear chains to higher dimensional sparse networks.
Finally, we present results of GNR networks for varying network density and size (network base area).
In \sref{sec:NA} we first introduce the NA model adjusted for our GNR quantum network purpose.
Then we present results of GNR networks with identical geometric parameters as for the QT calculations and compare the two approaches. We conclude that a single NA model cannot replicate all QT features, summarize our results and end with an outlook 
in \sref{sec:summary}.


\section{Quantum transport in nanoribbon networks}
\label{sec:QT}

\subsection{Model System}
\label{sec:QT:model_system}

\begin{figure}[!bt]
    \centering
    \includegraphics[width=0.9\textwidth]{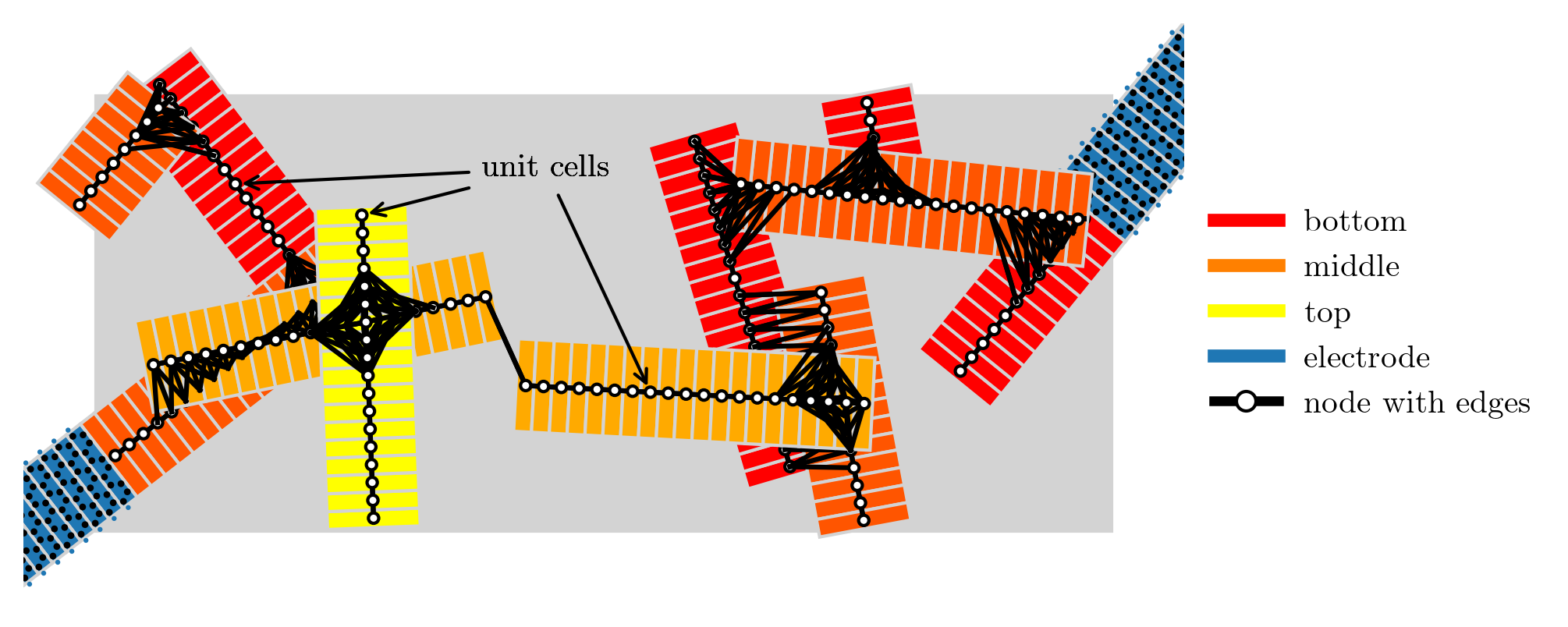}
    \caption{Model system of a QT network with $\varrho = 0.76$. The grey background marks the base area of $14 \times 6$ nm$^2$. The coloured stripes represent GNRs, which are divided into GNR unit cells. The black lines represent interactions between cells. The colour of the ribbons corresponds to their $z$-coordinate, ranging from red (lowest) to yellow (highest). Blue cells mark the semi-infinite electrodes where the alignment of the carbon atoms (black) and the attached hydrogen atoms is indicated.  
    }
    \label{fig:QT:network}
\end{figure}

The model system we study are GNR networks, see \fref{fig:QT:network}, which are constructed as follows.
A given number $N$ of GNRs are randomly distributed within the network base area $A$ by randomly selecting a point and a direction.
Here, we focus on one type of nanoribbon: The networks consist of zigzag GNRs with a width of 3 unit cells (also called 6-zGNR), which equals 12 carbon atoms (corresponding to a width of $1.3\,\mathrm{nm}$).
We use this type of GNR because it is large enough to describe realistic devices 
yet sufficiently small 
to obtain fast results.
Each GNR has a length of 20 unit cells, which equals $5\,\mathrm{nm}$. 
They are passivated to avoid unphysical dangling bonds, i.e.\ a hydrogen passivation is used as this occurs during the fabrication process~\cite{JAmChemSoc.135.2060}.
Thick strips in \fref{fig:QT:network} represent a single GNR (color coded for its spatial position), each of its small sub-stripes marks one unit cell. 
Each new GNR is placed in the lowest possible layer.
This means that if it would intersect with an GNR already placed, it is placed at the next higher layer with a vertical distance of $3.35\,\mathrm{\AA}$ (red / orange / yellow in \fref{fig:QT:network}) as this corresponds to the distance between two graphene layers~\cite{PhysicaE.40.3055}.
That means, the GNRs are assumed to be flat, and curvature effects are neglected.
In reality, overhanging ends of stacked GNRs could bend. 
The resulting curvature would lead to a reduced in-plane transmission and thus a reduced conductance.
This effect is in the range of up to 40\% reduction, depending on the curvature~\cite{ApplPhysLett.96.173101}.
Furthermore, additional connections between GNRs could occur, which would effectively lead to a larger network density $\rho$, thus counteracting the bending effect.
Curvature effects are not included 
here, not least to keep the model simple.

At the edges of the network base area, protruding 
GNR unit cells are shifted to the opposite side of the network imposing periodic boundary conditions.
For this system, the dimensionless network density $\varrho$ is defined as
\begin{equation}
    \varrho = N \Omega / A
\end{equation}
with the area covered by a single nanoribbon $\Omega = 6.39~\mathrm{nm}^2$ (for 6-zGNRs) and network area $A$.
For example, $\varrho=1.5$ means that the GNRs could fill one and a half layers if aligned properly.
Furthermore, one ribbon at the left and right network boundary is elongated infinitely to act as an electrode within the transport formalism (indicated in blue in \fref{fig:QT:network}).

\subsection{Quantum transport theory}
\label{sec:QT:quantum_transport_theory}

\begin{figure}[!tb]
    \centering
    \includegraphics[width=0.8\textwidth]{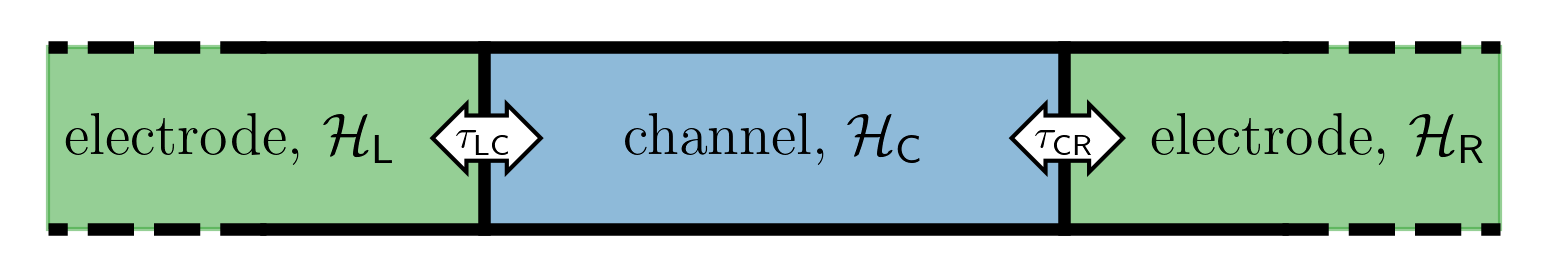}
    \caption{
    Schematic QT model system. Two electrodes (left L and right R respectively) are connected via a channel (C). Each component is represented by a Hamiltonian matrix $\mathcal{H}_\mathrm{L,C,R}$.}
    \label{fig:QT:device_scheme}
\end{figure}

Electronic transport is described quantum mechanically by using quantum transport theory~\cite{Datta2005} in combination with an underlying electron structure theory.
The conductance in the limit of a small bias can be computed from the transmission based on the Landauer-B\"uttiker formula~\cite{PhysRevB.31.6207}
\begin{equation} 
    G = - G_0 \int_{-\infty}^{\infty} T(E) \, \frac{\partial f(E)}{\partial E} \, \mathrm{d}E \quad.
\end{equation}
Here, $G_0=2\mathrm{e}^2/\mathrm{h}$ is the conductance quantum, $T(E)$ the transmission function, and $f(E)$ the Fermi distribution function.
In order to calculate the transmission function, the system is divided into a finite channel (C) and two semi-periodic electrodes (L -- left and R -- right) as depicted in \fref{fig:QT:device_scheme}.
Each part is described by a Hamiltonian matrix $\mathcal{H}_\mathrm{L,C,R}$ and coupling matrices $\tau_\mathrm{LC,CR}$. 
The resulting channel Green's function is
\begin{equation} 
    \GC = \left[ (E + \i \eta ) \mathcal{S}_\mathrm{C} - \HC - \SL - \SR \right]^{-1} \quad,
    \label{eq:QT:channel_greens_function}
\end{equation}
where $\mathcal{S}_\mathrm{C}$ is the channel overlap matrix, $\mathcal{S}_\mathrm{LC, CR}$   
are the coupling matrices of the overlap, and $\SL=(\tau_\mathrm{CL}-E\mathcal{S}_\mathrm{CL})\GL(\tau_\mathrm{LC}-E\mathcal{S}_\mathrm{LC})$ and $\SR=(\tau_\mathrm{CR}-E\mathcal{S}_\mathrm{CR})\GR(\tau_\mathrm{RC}-E\mathcal{S}_\mathrm{RC})$ are the self energy corrections due to the coupling of C to the electrodes L and R. Finally, $\eta$ is a small numerical value for improving convergence, which shifts the singularities at eigenenergies into the complex plane. 

The electrode surface Green's functions $\mathcal{G}_\mathrm{L/R}$ are calculated efficiently using the renormalization decimation algorithm (RDA), which is a fast recursive algorithm~\cite{JPhysFMetPhys.14.1205,JPhysFMetPhys.15.851}. We use a value of $\eta=10^{-5}$ for calculating $\GC$ and for $\mathcal{G}_\mathrm{L,R}$ via the RDA.

Finally, the transmission function is
\begin{equation} 
    T(E) = \Tr(\GaR \GC \GaL \GC^\dagger)\quad,
\end{equation}
where $\Gamma_\mathrm{L,R}=\rmi\left(\Sigma_\mathrm{L,R}-\Sigma_\mathrm{L,R}^\dagger\right)$ describe the broadening of the original channel states due to the coupling to the electrodes.

The Hamiltonian and coupling matrices themselves are computed in advance by an electronic structure calculation.
To treat the large network systems studied here, we use the so-called non-self-consistent DFTB model~\cite{PhysRevB.51.12947,IntJQuantumChem.58.185}, which readily provides the distance- and orbital-dependent Hamiltonian and overlap matrix elements.
For describing the organic molecules present here, the parameter set 3ob~\cite{JChemTheoryComput.9.338,PhysRevB.58.7260} is used.
This set uses a sp$^3$-basis appropriate for our purpose and has been verified against other organic molecules, especially sp$^2$-hybridised ones, with a similar structure.
The cutoff distance was chosen to be $a_\mathrm{CO} = 3.7$ \AA.
This value includes the 3rd nearest-neighbor interaction in plane but excludes the 4th one,
implying rather small GNR unit cells and reducing 
the decimation calculation time to an optimum.
In addition, it ensures the inclusion of 2nd nearest-neighbor interaction out of plane, i.e.\ between two GNRs in different layers.

\subsection{Network Transport: Reducing the Complexity through the Network Decimation Scheme}
\label{sec:QT:application}

Using the aforementioned methods, the conductance for an arbitrary system can in principle be obtained for a given, adequate parameter set.
However, due to the matrix inversion in \eref{eq:QT:channel_greens_function} the number of necessary operations scales cubic with the number of atoms in the system. 
This is alleviated by dividing the channel into smaller cells to calculate an effective channel Green's function $\tilde{\mathcal{G}}_\mathrm{C}$ that is equivalent to $\GC$ in \eref{eq:QT:channel_greens_function} but of lower dimension.

\begin{figure}[!b]
    \centering
    \begin{subfigure}[t]{.24\textwidth}
        \includegraphics[width=1.\textwidth]{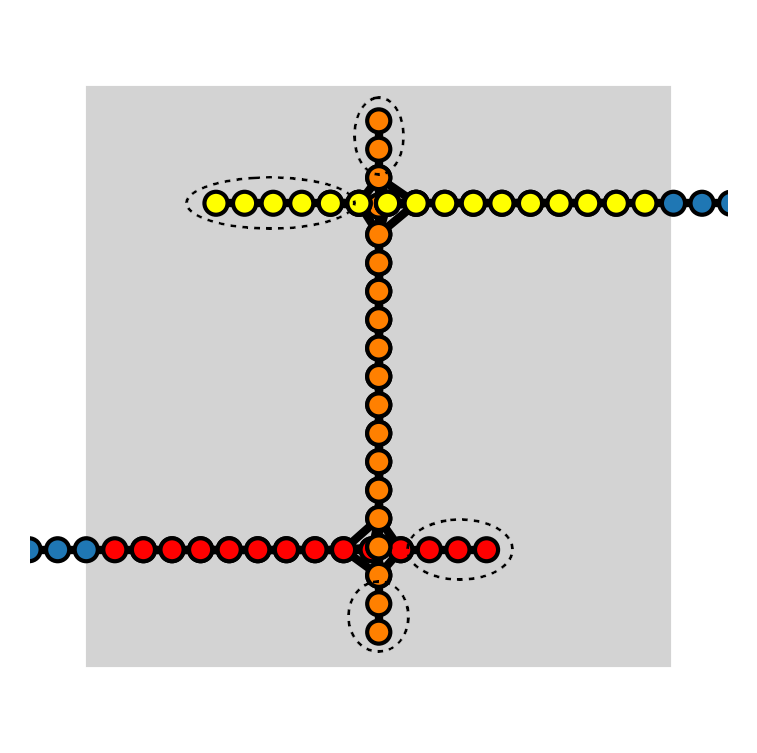}
        \caption{}
        \label{fig:QT:nw_decimation_a}
    \end{subfigure}
    \begin{subfigure}[t]{.24\textwidth}
        \includegraphics[width=1.\textwidth]{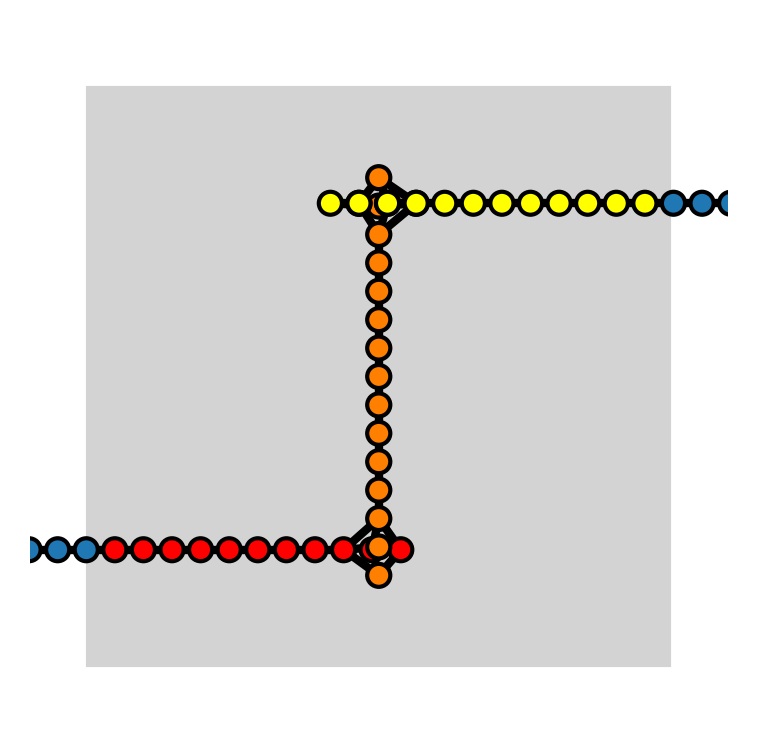}
        \caption{}
        \label{fig:QT:nw_decimation_b}
    \end{subfigure}
    \begin{subfigure}[t]{.24\textwidth}
        \includegraphics[width=1.\textwidth]{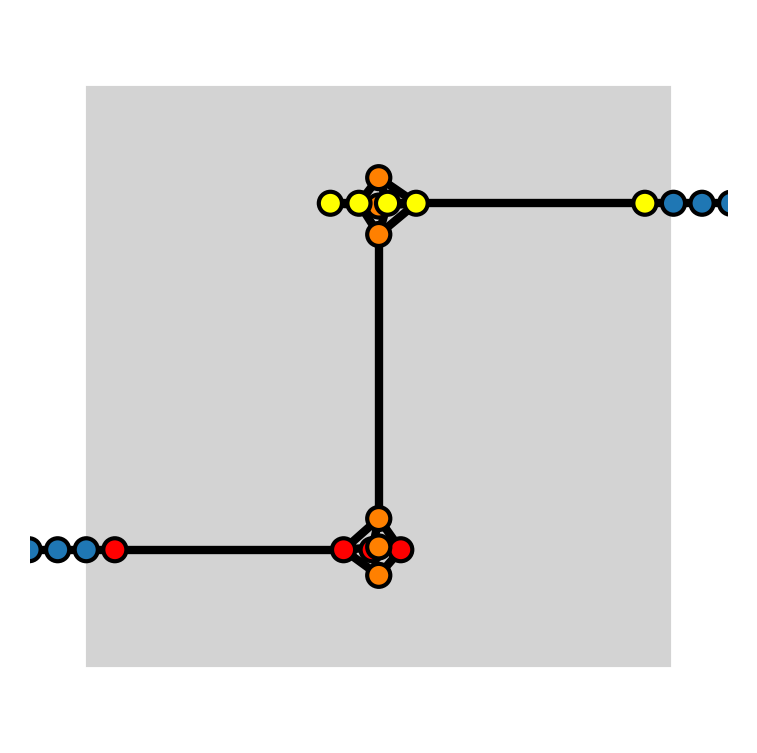}
        \caption{}
        \label{fig:QT:nw_decimation_c}
    \end{subfigure}
    \begin{subfigure}[t]{.24\textwidth}
        \includegraphics[width=1.\textwidth]{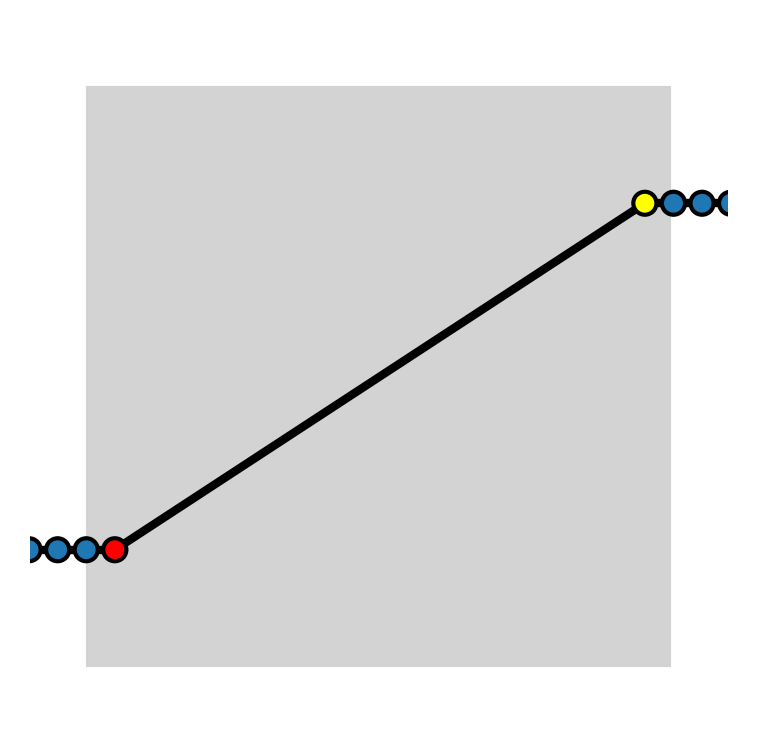}
        \caption{}
        \label{fig:QT:nw_decimation_d}
    \end{subfigure}
    \caption{
    Network decimation steps illustrated in the sequence (a) through (d). The coloured dots represent nodes which are connected by edges (black lines). The colour corresponds to the $z$-coordinate of the cell, as described in \sref{sec:QT:model_system}. The blue dots represent the semi-infinite electrodes.
    See text for details.
    }
    \label{fig:QT:nw_decimation}
\end{figure}

Proficient algorithms benefit from this subdivision and result in a linear scaling with the number of atoms.
For example, the recursive Green's function formalism (RGF)~\cite{JPhysCSolidStatePhys.14.235,ZPhysBCondMat.59.385} has been widely used for similar calculations of linear chains~\cite{cresti2008charge, NanoLett.9.2725, JPhysCondMat.26.045303, PhysStatSolB.247.2962, NJPhys.16.123026, ComputMatSci.138.49, JPhysCommun.2.105012}.
Two slightly different versions of the RGF, the renormalization decimation scheme (RDS) and forward iteration scheme (FIS) are utilized~\cite{JComputPhys.334.607} and adopted to the networks studied here, as shown in \fref{fig:QT:nw_decimation}.

These methods are combined in the network decimation scheme introduced here to reduce the complexity of the network incrementally by virtually decreasing the number of nodes that need to be considered. The starting point is a system where each GNR unit cell, cf.~\fref{fig:QT:network}, corresponds to one node. 

The RDS is given by the equations 
\begin{eqnarray}
\eqalign{
    \mathcal{G}_i &:= \left[(E + \i \eta) \mathcal{S}_i - \mathcal{H}_i \right]^{-1}\\
    \mathcal{H}_j &:= \mathcal{H}_j + \tau_{j,i} \mathcal{H}_i \tau_{i,j}\\
    \tau_{j,k} &:= \tau_{j,i} \mathcal{G}_i \tau_{i,k} \quad .
}\label{eq:QT:rds}
\end{eqnarray}
The indices $i,j,k$ refer to the respective nodes.
$\tau_{i,j}$ is the coupling matrix from node $i$ to node $j$.
The presence of a node causes a shift of all the neighbouring energy levels. 
Using RDS, this shift is applied to the Hamiltonian and coupling matrices of the surrounding nodes.
Thus, after applying \eref{eq:QT:rds} node $i$ is effectively eliminated as all its information is now stored in its neighbours and the connections between them.

The FIS takes advantage of the fact that the overall Hamiltonian is sparse and that the electrodes only couple to a small part of the system.
It is defined by
\begin{eqnarray}
\eqalign{
    \mathcal{G}_1 &:= \left[(E + \i \eta) \mathcal{S}_1 - \mathcal{H}_1 - \Sigma_\mathrm{L}\right]^{-1} \\
    \mathcal{G}_i &:= \left[(E + \i \eta) \mathcal{S}_i - \mathcal{H}_i - \tau_{i,i-1} \mathcal{G}_{i-1} \tau_{i-1,i}\right]^{-1} \\
    \mathcal{G}_N &:= \left[(E + \i \eta) \mathcal{S}_N - \mathcal{H}_N - \tau_{N,N-1} \mathcal{G}_{N-1} \tau_{N-1,N} - \Sigma_\mathrm{R} \right]^{-1} \\
    \tilde{\mathcal{G}}_\mathrm{C} &:= \mathcal{G}_N \times \tau_{N,N-1} \times \mathcal{G}_{N-1} \times ... \times \tau_{2,1} \times \mathcal{G}_1 \quad ,
}\label{eq:QT:fis}
\end{eqnarray}
which represents the influence of the electrodes and the neighbourhoods of the cells propagating through a linear chain of cells.
In the end $\tilde{\mathcal{G}}_\mathrm{C}$ is obtained, which can be understood as an effective Green's function of the system.

As a network of cells as shown in \fref{fig:QT:network} is far from constituting a linear chain, \eref{eq:QT:rds} and \eref{eq:QT:fis} are combined to obtain $\tilde{\mathcal{G}}_\mathrm{C}$.
In order to minimise the number of computer operations, they are applied as follows: 
We start the decimation scheme at the outermost edges, where the nodes have only one neighbour (see circled cells in \fref{fig:QT:nw_decimation_a}).
These cells are successively decimated using RDS, until only clusters and connections between them remain (see \fref{fig:QT:nw_decimation_b}). 
Then, connections between clusters are treated, where nodes only have two edges each.
Subsequently, the RDS is applied to the clusters of many interconnected cells, where the nodes with the fewest edges are decimated one at a time (see \fref{fig:QT:nw_decimation_c}). 
In the end, only the cells connected to the electrodes remain present (see \fref{fig:QT:nw_decimation_d}), as they can't be decimated using RDS. 
In this final step, FIS is used and $\mathcal{G}_\mathrm{C}$ is obtained. 
We implemented this QT algorithm in python using numpy and scipy.
For further details we would like to refer the interested reader to~\cite{Masterarbeit.Rodemund}.

This hierarchic decimation procedure works especially well for weakly connected networks, i.e. networks with few overlapping areas compared to the total network. 
The standard recursive procedure, which is often used for linear chains, would provide a subdivision in transport direction only.
For linear chains this results in an $\mathcal{O}(N)$ scaling with the number of atoms $N$.
For 2D and 3D networks, it yields a $\mathcal{O}(N^2)$ and $\mathcal{O}(N^{7/3})$, respectively, scaling, because only one dimension with $N^{1/d}$ atoms scales linear and the other dimensions with $N^{(d-1)/d}$ atoms scale cubical due to the matrix inversion.
The decimation scheme presented here treats this more efficiently as all directions are subject to decimation.
In the limit of weakly connected networks, this implies a vanishing fraction of overlapping areas (node clusters in \fref{fig:QT:nw_decimation}) compared to the original network, and the algorithm scales like $\mathcal{O}(N)$ independent of the network dimension.
The complexity scaling is summarised in \tref{tab:complexity}.

\begin{table}[!b]
    \begin{tabular}{l|c|c|c}
        & 1D & 2D & 3D \\
        \hline
        standard decimation & $\mathcal{O}(N)$ & $\mathcal{O}(N^2)$ & $\mathcal{O}(N^{7/3})$ \\
        \hline 
        network decimation (this work:) & & \\
        weakly connected networks & $\mathcal{O}(N)$ & $\mathcal{O}(N)$ & $\mathcal{O}(N)$ \\
        strongly connected networks & $\mathcal{O}(N)$ & $\mathcal{O}(N^2)$ &
    \end{tabular}
    \caption{Complexity scaling of the standard decimation method and our network decimation approach. $N$ is the total number of atoms in the system.}
    \label{tab:complexity}
\end{table}

However, such a $\mathcal{O}(N)$ scaling is not achievable for dense, realistic systems in general. Clusters with a high degree of interconnections scale with $\mathcal{O}(N^2)$ and contain typically between $50$ to $90 \%$ of node sites.
Nevertheless, in comparison to the standard procedure with a $N^2$ scaling for 2D systems, a scaling between $N$ and $N^2$ can be achieved using the novel network decimation scheme, which allows one to treat larger systems by  quantum transport methods, as the one we discuss here.
Of course, in the limit of denser networks, the linear scaling breaks down.
In the extreme limit where the whole network is one cluster, a quadratic scaling $\mathcal{O}(N^2)$ is obtained, which is the same as for 2D systems treated with the standard algorithm of linear chains.
For 3D systems, a slightly better scaling may be achievable, but this is out of the scope of our work as we focused on quasi 2D networks with few layers.

\subsection{Results}
\label{sec:QT:results}

For each set of geometric parameters (GNR type, area, number of GNRs) 500 random and percolating networks were generated, for which the QT calculations are performed and statistically analysed.
Percolating or non-percolating networks could arise due to the finite interaction distance given by the cutoff distance $a_{\mathrm{CO}}$.

\begin{figure}[!t]
    \centering
    \includegraphics[width=0.7\textwidth]{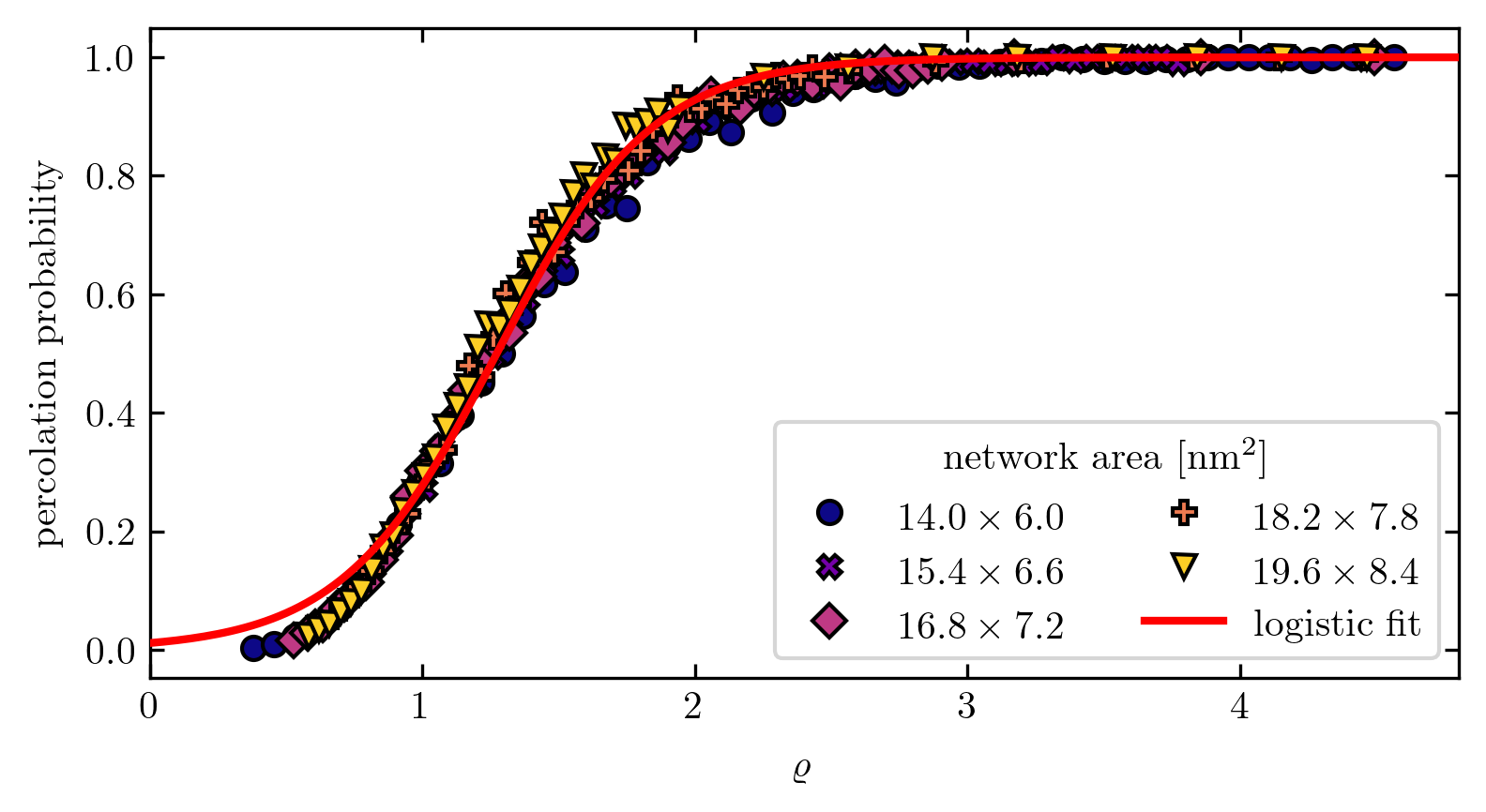}
    \caption{
    Percolation probability of QT systems with 6-zGNRs for different network sizes (base area as denoted by coloured symbols). The red line represents a logistic approximation of the data with $P(\varrho) = \left\{ 1 + \exp [ 0.28 \cdot (1.27 - \varrho)] \right\}^{-1}$. 
    }
    \label{fig:QT:percolation}
\end{figure}

We first state that the usual percolation behaviour is observed:
\Fref{fig:QT:percolation} shows the percolation probability as a function of the network density for different network sizes (base areas $A$ denoted by colour).
There is a minimal network density below which the percolation probability is exactly zero because the cumulative length of the GNRs is smaller than the side length of the network base area.
At the percolation threshold network density the percolation probability increases markedly and tends towards one for large densities.
The overall trend can be approximately described by a logistic function fit (red line in \fref{fig:QT:percolation}) \cite{doi:10.1021/j150299a014}. A finite-size effect can be seen as larger network base areas result in steeper curves, that is, sharper transitions. However, percolation aspects are not the focus of this work, and we refer the reader to \cite{shante1971introduction, essam1980percolation}, while we focus on percolating network ensembles in the following. 
From an engineering point of view, non-percolating networks would be ``broken devices'' and not relevant for applications.

\begin{figure}[!tb]
    \centering
    \includegraphics[width=0.7\textwidth]{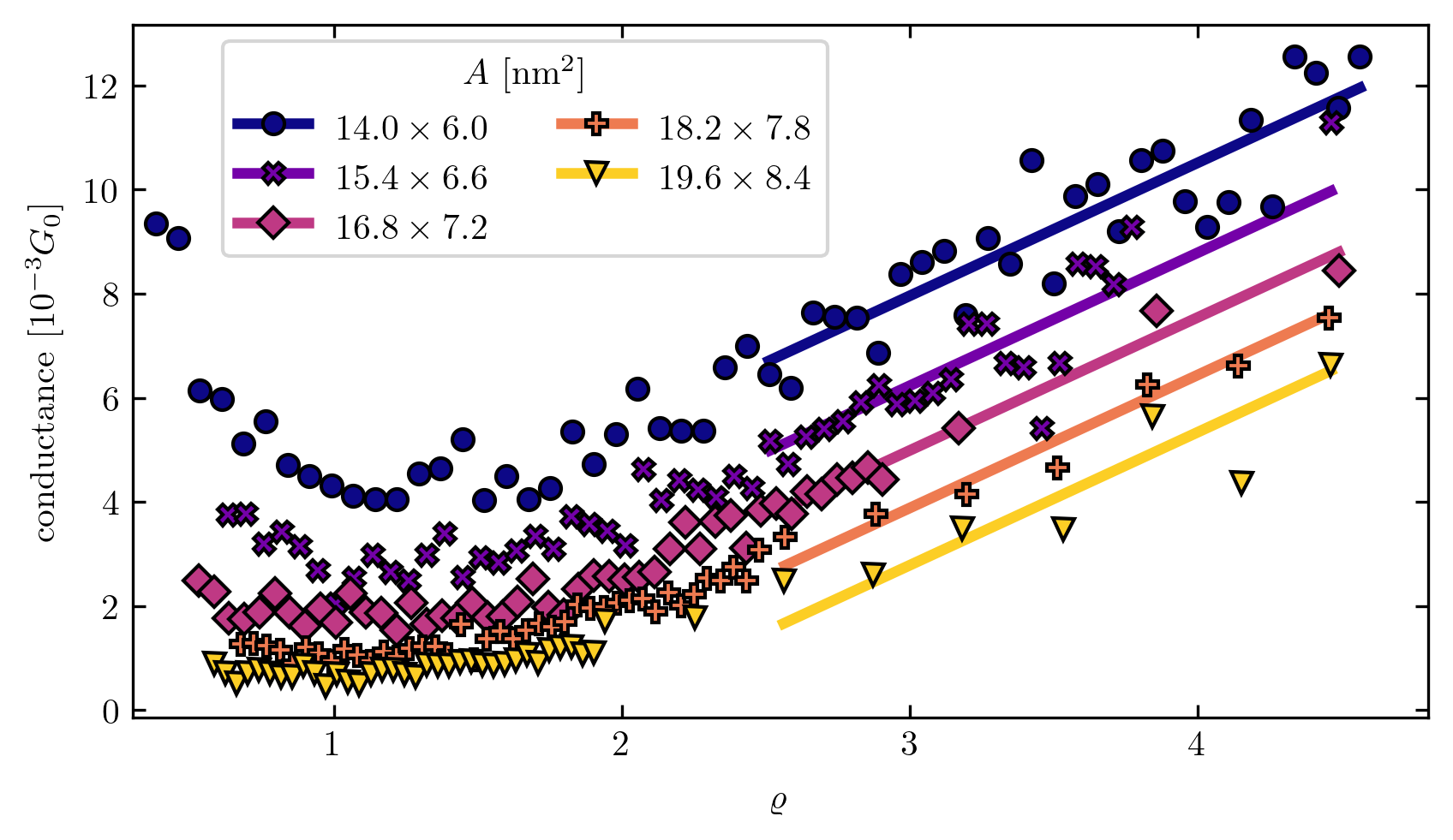}
    \caption{
    Conductance of percolating quantum networks depending on the network density $\varrho$ for different network sizes (base area $A$ denoted by coloured symbols, same as in \fref{fig:QT:percolation})
    Two quantum transport mechanisms, the linear chain regime for $\varrho <1$ and the diffusive regime with linear increase (indicated by the solid, fitted lines) for  $\varrho > 1$ can be distinguished.
    }
    \label{fig:QT:conductance}
\end{figure}

\Fref{fig:QT:conductance} shows the average conductance of percolating networks as a function of the network density for various network sizes (base area $A$ denoted by colour as before).
For each data set (i.e., for a fixed base area $A$), two different regimes can cleraly be distinguished:

(A) The ``linear chain regime'' for low densities $\varrho<1$.
Here, the networks are weakly connected and possess few, often only one transport path.
They thus behave like an effectively linear chain with few perturbations.
In the limit of a network consisting only of one GNR covering the whole system (not included in \fref{fig:QT:conductance}), this behaviour tends towards the ballistic transport regime with constant conductance $G=2G_0$.
In a macroscopic picture for larger network sizes, this can be interpreted as a series circuit, where the conductance decreases roughly like $G\sim 1/N$ with $N$ being the number of tunnelling regions between any two GNRs contribution to the chain.
$N$ increases with increasing effective chain length $\ell_\mathrm{eff}$, which in turn increases with density $\varrho$ as long as only one chain exists:  $N\sim\ell_\mathrm{eff}\sim\varrho$.
Thus, the conductance decreases with increasing density $G\sim 1/\varrho$, as can be seen in \fref{fig:QT:conductance}, especially for small base areas $A$.

(B) The ``interference dominated diffusive regime'' for high densities $\varrho>1$.
Here, the networks are highly connected, i.e. a significant amount of the networks contributes to percolation and many transport paths are present.
An increase of the network density results in further increasing the number of connections and transport paths, thereby increasing the conductance.
In a macroscopic picture this can be interpreted as a parallel circuit, where the conductance increases roughly like $G\sim M$ with $M$ being the number of parallel GNR paths.
$M$ increases with increasing number of layers or system height $H$, which in turn increases with increasing network density $\varrho$, leading to $M\sim H\sim\varrho$.
Thus, the conductance increases with increasing density $G\sim \varrho$, as can be seen in \fref{fig:QT:conductance} for all base areas $A$.
A linear regression for $\varrho\geq 2.5$, i.e. the range where virtually all networks are percolating, has been done (solid lines in \fref{fig:QT:conductance}), yielding a slope
\begin{eqnarray}
    \frac{\partial G}{\partial \varrho} &= 2.55 \cdot 10^{-3} \, G_0 \quad.
    \label{eq:qt:linfits}
\end{eqnarray}

Both trends (A) and (B) can be seen best for the smaller network base areas, e.g. the system with a base area of $14 \times 6 \, \mathrm{nm}^2$.
In between both regimes (A) and (B), a transition region must be present and is realised as a plateau-like minimum around $\varrho \approx 1$.

In addition, the conductance for a fixed density $\varrho$ depends on the network size (base area $A$ indicated by colours as in \fref{fig:QT:conductance}).
To this end we increased the area $A=WL$ in width $W$ and length $L$, while fixing the aspect ratio $W/L=\mathrm{const}$.
In the ``linear chain regime'' (A) increasing $A$ leads also to an increased effective chain length $\ell_\mathrm{eff}$ and thus a decreased conductance, as can be seen in \fref{fig:QT:conductance}.
In the ``interference dominated diffusion regime'' (B)  increasing $A$ leads on the one hand to an increased number of parallel paths due to the increased width $W$, which will increase the conductance $G\sim W$.
On the other hand the increased length requires more tunnelling events between two GNRs.
But when the system gets very large the scattering events in the diffusive regime of 1D systems lead to enhanced interference and strong localisation effects, which will decrease the conductance exponentially with length $L$.
In the mesoscopic range of the networks discussed here, the strong localisation regime may not be fully reached yet (as shown for similar purely 1D systems \cite{NJPhys.16.123026, ComputMatSci.138.49, JPhysCommun.2.105012}).
In summary, we get $G\sim H \, W/\mathrm{e}^L$, which explains the increasing conductance with increasing $\varrho\sim H$ and the decreasing conductance with equally increasing $W$ and $L$ (i.e. increasing $A$).


\section{Comparison with semi-classical transport and nodal analysis}
\label{sec:NA}

In this chapter we will address the question to what extent our results for GNR networks could be reproduced based on a classical transport approach. Quantum-classical correspondence has been established as a useful tool in mesoscopic physics. The classical approach neglects all interference effects, and comparison with the quantum transport results allows one thus to classify their importance and a deeper characterisation of the system properties. 

\subsection{Model System}
\label{sec:NA:model_system}

\begin{figure}[!tb]
    \centering
    \includegraphics[width=0.9\textwidth]{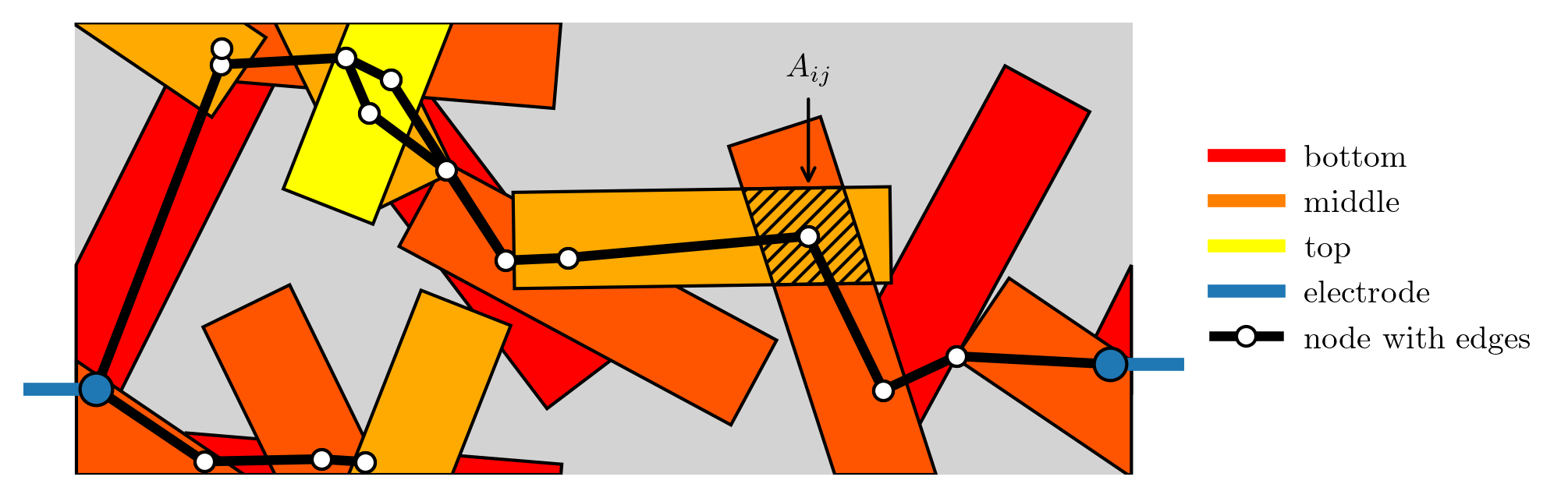}
    \caption{
    NA network with $\varrho = 0.76$. A network consisting of GNRs is contacted by two electrodes, which are connected to one another via a system of nodes and edges. The dashed area marks exemplarily the overlapping area of two ribbons used for the calculation of $\condmat^\mathrm{inter}_{ij}$ (see \eref{eq:NA:Ginter}).
    }
    \label{fig:NA:model_system}
\end{figure}

In order to approach the GNR networks in \sref{sec:QT} using classical transport based on nodal analysis (NA), the networks are represented by layers of two-dimensional (2D) polygons (see \fref{fig:NA:model_system}). 
Each polygon has a width and length corresponding to their atomic (quantum) counterpart, which is roughly $1.3 \times 5~\mathrm{nm}^2$ in the 6-zGNR case. 
These stripes are now randomly placed on the base area $A = L W$, starting in the lowest layer. 
Analogous to \sref{sec:QT:model_system}, overhanging stripes are shifted to the opposing side of the base area.

Two stripes are deemed interacting, if they are in adjacent layers and overlap in the $x,y$ plane. 
The centre of the overlapping region determines the $x,y$ coordinates of the two resulting nodes, one on each ribbon taking part in the interaction. 
These coordinates are then used to order the emerging nodes of a ribbon, so that each node is connected to its nearest neighbours. 
This way the network consisting of ribbons is represented as nodes and edges, which are then investigated using classical NA.

Using Kirchhoff's current law, an arbitrary circuit consisting of ohmic resistances can be expressed as
\begin{equation}
    \condmat \bvarphi = \bi{I} \quad ,
\end{equation}
where the conductance matrix $\condmat$ represents the connections between the nodes, the vector $\bvarphi$ indicates the potential at each node and $\bi{I}$ contains the net currents in each node. 
The currents are all zero due to Kirchhoff's first law, with the exception of two contacts where the current is inserted into or extracted from the system (see blue markers in \fref{fig:NA:model_system}).

$\condmat$ must be chosen to best reflect the results obtained for the quantum system in \sref{sec:QT}. Thus, the following definition is used:
\begin{equation}\label{eq:NA:condmat}
    \condmat_{ij} = \left\{
    \begin{array}{ll}
        \condmat^\mathrm{intra}    & \textrm{nodes $i,j$ on the same ribbon} \\
        \condmat^\mathrm{inter}_{ij}   & \textrm{nodes $i,j$ on adjacent ribbons} \\
        - \displaystyle\sum_{k \neq i} \condmat_{kj}    & i = j \\
        0                           & \textrm{nodes not connected}
    \end{array}\right. \quad .
\end{equation}
The choice for the $j=k$ and the unconnected node case
is motivated by the NA definition. 
Additionally, one needs to distinguish between intra-ribbon connections and inter-ribbon connections. 
The conductance between two nodes on the same ribbon is assumed to be $G^\mathrm{intra} = 2 G_0$, which corresponds to the conductance of an ideal GNR with ballistic transport. 
For two nodes on separate ribbons
\begin{equation}\label{eq:NA:Ginter}
    \condmat^\mathrm{inter}_{ij} = g_\mathrm{t} A_{ij}
\end{equation}
is chosen, effectively describing electron tunnelling. 
This conductance is assumed proportional to the overlapping area $A_{ij} = A_{ji}$ of the ribbons between two nodes $i$ and $j$ (see \fref{fig:NA:model_system}). 
The factor $g_\mathrm{t}$ is to be optimized as to yield the best agreement with the quantum transport results obtained in the previous \sref{sec:QT}.

\subsection{Results}
\label{sec:NA:results}

\begin{figure}[!tb]
    \centering
    \includegraphics[width=0.6\textwidth]{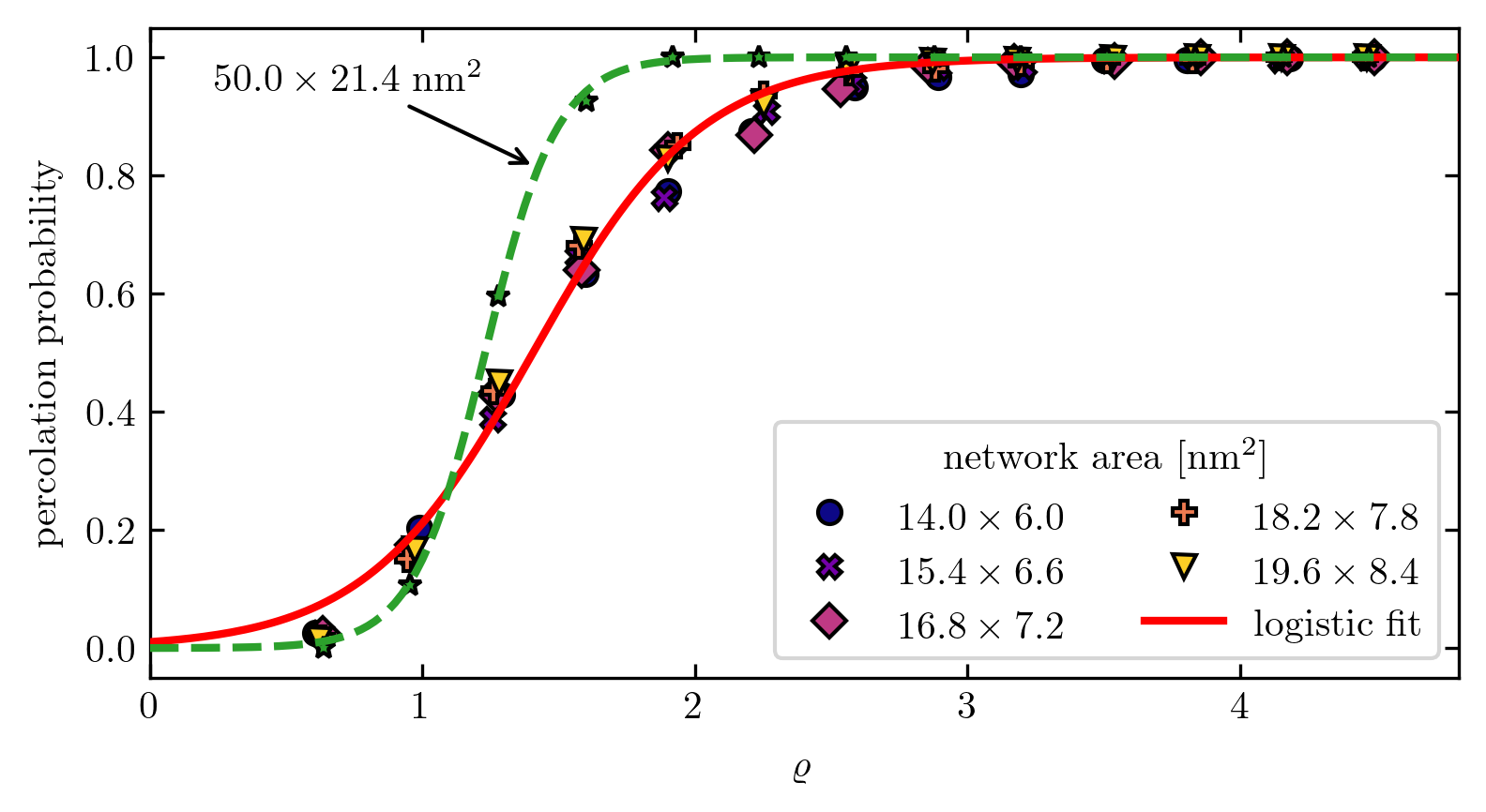}
    \caption{
    Percolation probability of NA networks of various base area sizes. The logistic fit (red line) yields $P(\varrho) = \left\{ 1 + \exp [ 0.31 \cdot (1.41 - \varrho)] \right\}^{-1}$.
    }
    \label{fig:NA:percolation_probability}
\end{figure}

\Fref{fig:NA:percolation_probability} depicts the percolation probability of the NA approach for different network base areas.
The qualitative behaviour is the same as described in \sref{sec:QT:results}.
Quantitatively there are some small differences in the threshold density and slope in comparison to the QT results that can be related to the differences between the models used. 

\begin{figure}[!b]
    \centering
    \includegraphics[width=0.6\textwidth]{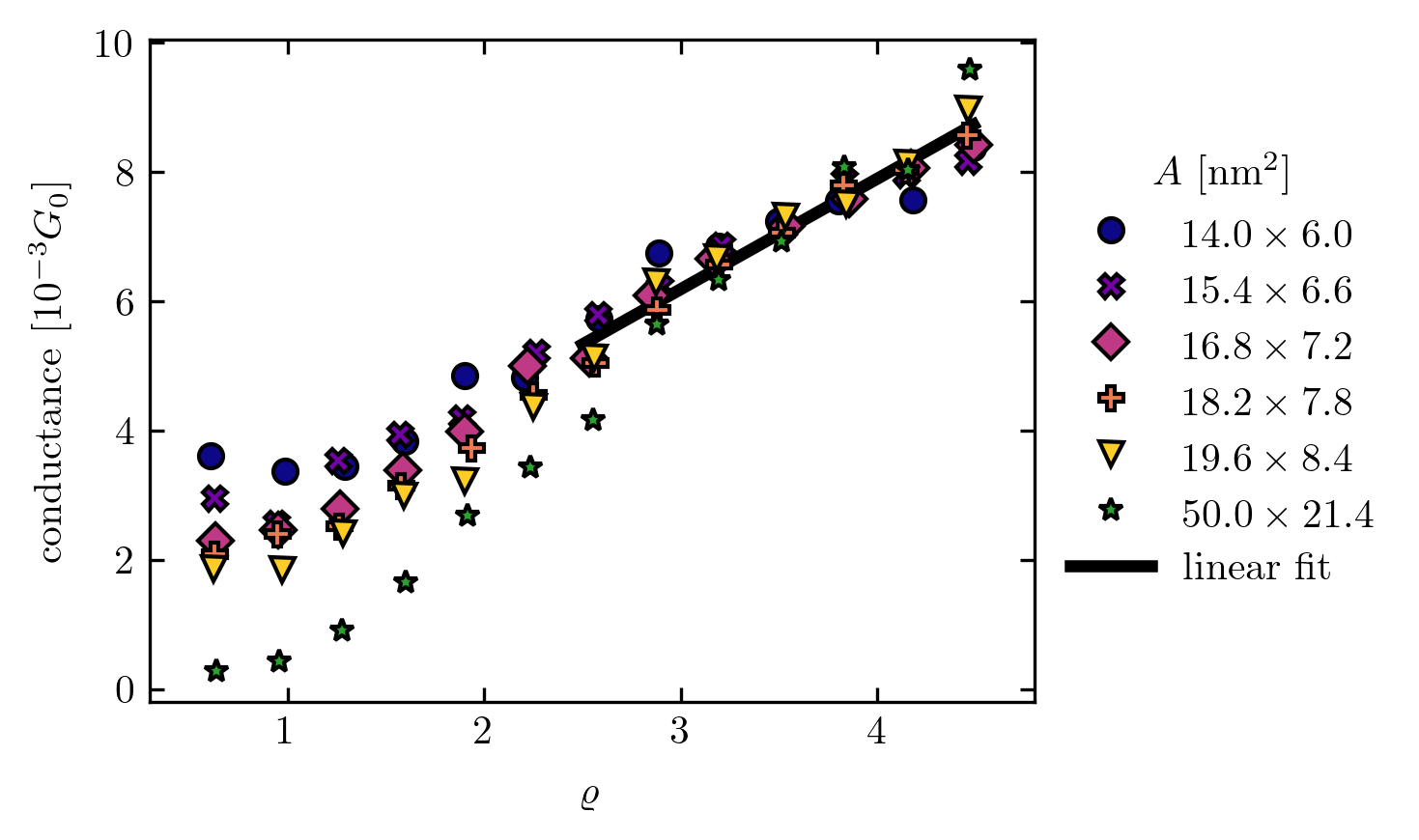}
    \caption{Conductance of 2D-NA networks for $g_\mathrm{t} = 0.1 \, G_0 / \mathrm{nm}^2$.}
    \label{fig:NA:conductance}
\end{figure}

\Fref{fig:NA:conductance} shows the conductance as a function of the network density for different network base areas (denoted by colour), calculated with the NA approach.
Analogous to the QT results in \sref{sec:QT:results}, two regimes can be found for each data set (fixed base area).

(A) The ``linear chain regime'' for low densities $\varrho<1$.
The general qualitative trends are similar to the ones of the QT
results in \fref{fig:QT:conductance}:
Weakly connected networks in the limit of only one chain yield a conductance $G \sim 1/L \sim 1/\varrho$ that decreases with increasing length and density, which in a macroscopic interpretation corresponds to a series circuit.
The decrease with density $\varrho$ and base area $A$ can be seen in the diagram.
The effect is less pronounced than for the quantum results.
Nevertheless, the limit of only one lead spanning the whole system $G = 2 G_0$ is the same (not shown).

(B) The ``ohmic regime'' for high densities $\varrho>1$.
Also here, the trend for fixed base area is similar to the quantum transport calculations.
Highly connected networks with many transport paths yield an increasing conductance $G \sim H \sim \varrho$ with increasing height and density, as can be seen in the diagram.
In a macroscopic interpretation this corresponds to a parallel circuit.
In contrast to the quantum treatment before, however, the base area $A$ dependence is now different:
Although the conductance increases with width $W$ as before,
the length dependence differs because interference effects and thus an exponential conductance reduction with increasing length cannot be captured in NA calculations.
Thus, a (only) roughly inverse length dependence $G \sim 1/L$ is found.
In total, we find the Ohmic behaviour $G \sim HW/L$.
As we kept the aspect ratio $W/L$ constant for all $A$,
the conductance $G$ depends only on height $H$ and not on area $A$.
This can nicely be seen in \fref{fig:NA:conductance} as all curves roughly fall together.
The solid line represents a linear regression in this ohmic regime for $\varrho \geq 2.5$ where all networks are percolating.
The corresponding derivative yields the slope
\begin{eqnarray}
    \frac{\partial G}{\partial \varrho} &= 1.71 \cdot 10^{-3} \, G_0
    \label{eq:na:linfits}
\end{eqnarray}
for the specific parameter $g_\mathrm{t}=0.1\,G_0/\mathrm{nm}^2$.
As the derivative depends on $g_\mathrm{t}$, it could be adjusted such that the NA value equals the QT value, however, only for this specific 
value $g_\mathrm{t}$. To this end one could 
perform multiple NA calculations with many different $g_\mathrm{t}$ and choosing 
the best fitting $g_\mathrm{t}$ by minimising the difference between the QT and NA 
conductances.
Nevertheless, this $g_\mathrm{t}$  will be specific for the 
set of geometric parameters chosen.
A generic value of $g_\mathrm{t}$, that brings the NA calculations in agreement with QT calculations for arbitrary geometries, cannot be found.

A comparison 
of the density dependence of the conductance $G(\varrho)$ in the quantum transport (QT) and nodal analysis (NA) model reveals a semi-quantitative agreement in the behaviour for low $\varrho$ (linear chain regime). However, for high $\varrho$ characteristic deviations are visible, the most striking being that the conductance $G$ does not depend on the network size (or base area $A$) in the NA model due to the ohmic behavior. In contrast, quantum interference effects depending on the network size $A$ change this behavior in the QT model and result in an $A$ dependence that cannot be reproduced with the NA model. However, the NA parameter $g_\mathrm{t}$ can be tuned such that a best fit to the QT case is obtained for a (single) given network of base area $A$. 

\begin{figure}[!tb]
    \centering
    \includegraphics[width=0.6\textwidth]{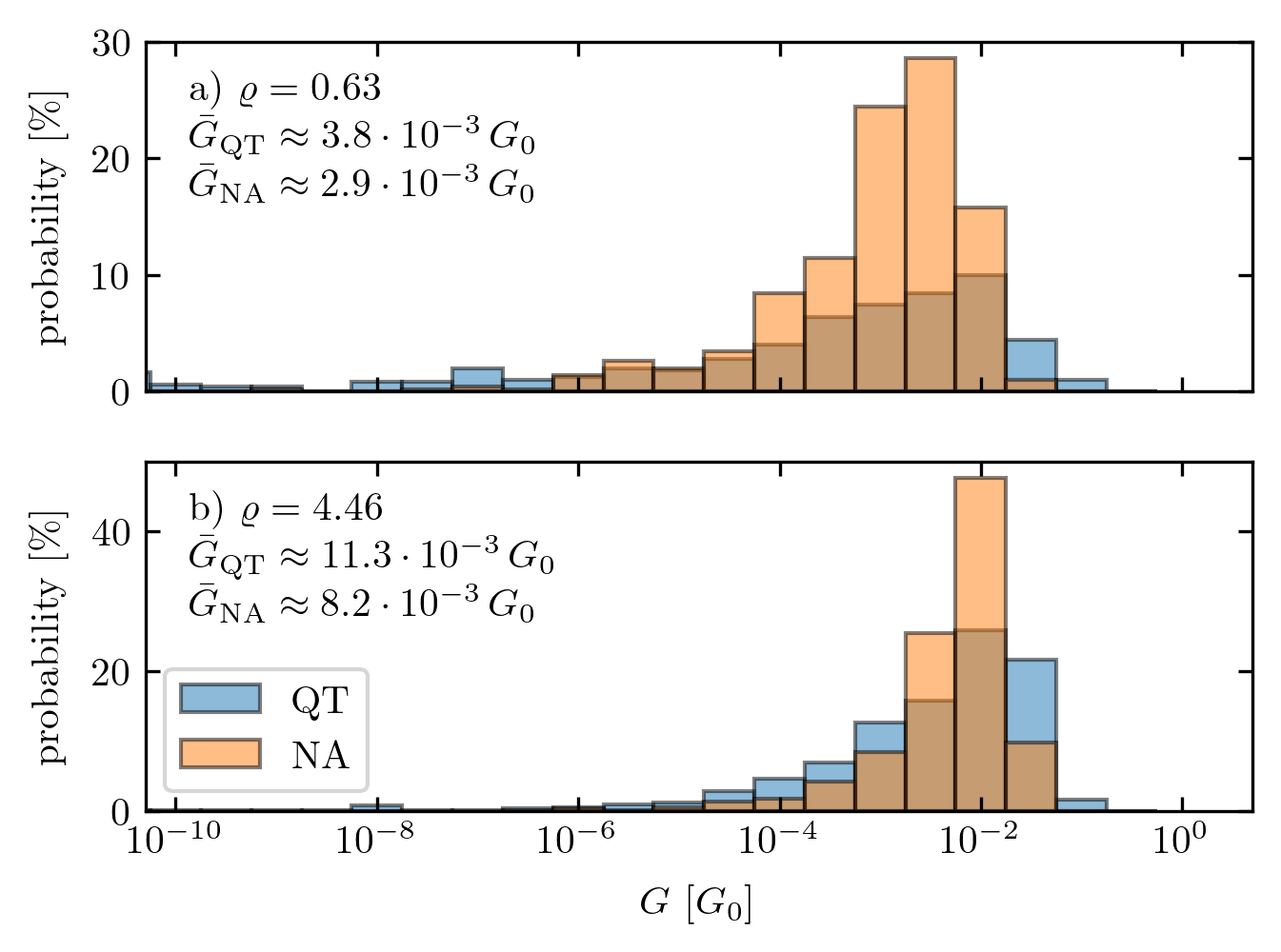}
    \caption{Conductance distributions for a) the linear chain regime at low densities and b) the interference-dominated diffusion regime at high densities. For the NA calculations, $g_\mathrm{inter} = 0.1 \, G_0 / \mathrm{nm}^2$ was used.
    }
    \label{fig:comp:conductance_distribution}
\end{figure}

Finally, we briefly discuss the distribution of the conductance within a statistical analysis.
This is shown in \fref{fig:comp:conductance_distribution} for QT (blue) and NA (orange) within a) the low and b) the high density regime.
Comparing both regimes, the distribution is much broader for low $\varrho$ than for high $\varrho$.
This is due to the fact that in the linear chain regime (low $\varrho$) the conductance depends sensitively on the specific tunnelling geometry. 
For the large amount of geometries where the GNRs are barely touching each other the conductance is fully determined by the corresponding tunnelling distance.
For the NA there is a hard cutoff and thus no such long tail is present.

In the high density regime, many transport paths exist within the network.
Consequently the electrons can choose between different pathways to avoid tunnelling-restricted geometries, which leads to a suppression of the low-conductance tail of the distribution in \fref{fig:comp:conductance_distribution}b) in comparison to \fref{fig:comp:conductance_distribution}a).
However, the QT calculations allow for strong localization effects as discussed before.
Thus, specific geometries show an exponentially suppressed conductance even in the high $\varrho$ situation. 
This explains the low conductance contributions for the QT model (blue) in \fref{fig:comp:conductance_distribution}b).


\section{Summary and Outlook}
\label{sec:summary}

The network decimation scheme introduced in this paper represents an efficient decimation scheme for quantum transport, which enables the treatment of comparably large quantum networks, that otherwise would not be accessible using standard recursive procedures or even direct inversion. The conductances of GNR networks with varying density were calculated using this novel algorithm and compared with results from nodal analysis as a classical transport method.
Although these two approaches may be able to yield similar results for specific geometries in the large-density regime for $\varrho \geq 2.5$ by parameter tuning, single parameterization of a nodal analysis model cannot generically reproduce the QT results with all features for different geometric parameters.
For low densities $\varrho$ quantum effects such as tunnelling and interference dominate and yield an increased conductance, which is only partly accessible by semi-classical nodal analysis.
The dependence on the network size (base area $A$) for high $\varrho$ observed for QT cannot be reproduced by NA.
In turn, the numerically and conceptually cheap nodal analysis can be expected to yield qualitatively reliable results for specific GNR networks with given, fixed geometric parameters and for $\varrho \geq 2.5$.
In summary, the efficient network decimation algorithm introduced here is both necessary to accurately 
calculate the conductivity of GNR-networks and allows one to extent full QT calculation to large systems that were previously only accessable by less correct semiclassical methods.  

The work presented here will be the basis for manifold future investigations ranging from accessing extended parameter regimes (aspect ratio of the network, GNR width, or GNR type) to other material systems (carbon nanotubes or other types of nanowires), and is not restricted to carbon.
The extension to other networks, for example the simulation of porous materials, is possible.
Eventually, special geometries can be investigated, such as periodically repeating structures, bent/curved structures or multi-terminal devices.


\end{document}

%% file: arxiv.tex
\KOMAoptions{paper=a4,paper=portrait,fontsize=10pt}
\usepackage[english]{babel}
\usepackage[ansinew]{inputenc}
\usepackage[T1]{fontenc}
\usepackage{helvet}
\usepackage[a4paper,portrait]{geometry}
\let\oldtwocolumn\twocolumn

\newif\iftwocolumn\twocolumntrue
\def\onecolumn{\twocolumnfalse}
\def\twocolumn{\twocolumntrue}
\usepackage{indentfirst}
\newlength{\OneColumnWidth}
\newlength{\TwoColumnWidth}
\makeatletter
\renewcommand\section{\scr@startsection{section}{1}{\z@}{-3.5ex \@plus -1ex \@minus -.2ex}{2.3ex \@plus.2ex}{\normalfont\bfseries}}
\renewcommand\subsection{\scr@startsection{subsection}{2}{\z@}{-3.5ex \@plus -1ex \@minus -.2ex}{2.3ex \@plus.2ex}{\normalfont\bfseries}}
\makeatother

\usepackage[
	colorlinks=true,
	urlcolor=blue, 
	filecolor=blue, 
	linkcolor=blue, 
	citecolor=blue, 
]{hyperref}
\usepackage{cite} 

\usepackage[singlelinecheck=true,font=small,labelfont=bf,format=plain]{caption} 

\let\oldhline\hline
\renewcommand{\hline}{\oldhline\rule{0pt}{12pt}}
\usepackage{enumitem}
\setlist{nosep}
\setenumerate[1]{label=(\arabic*)}
\setenumerate[2]{label=(\Alph*)}

\usepackage{ifthen}
\usepackage{forloop}
\usepackage{etoolbox}
\usepackage{graphicx}
\usepackage[fleqn]{amsmath}

\renewcommand{\title}[1]{\def\inserttitle{#1}}
\newcommand{\email}[1]{\def\insertemail{#1}}
\renewcommand{\abstract}[1]{\def\insertabstract{#1}}
\def\insertjournal{}
\def\insertdoi{}
\def\insertarxiv{}
\newcommand{\journal}[3][accepted]{
	\def\tmpa{#1}\def\tmpb{submitted}\ifx\tmpa\tmpb\def\journalpre{Submitted to: }\else\def\tmpb{accepted}\ifx\tmpa\tmpb\def\journalpre{Accepted for publication in: }\else\def\tmpb{prepared}\ifx\tmpa\tmpb\def\journalpre{Prepared for submission to: }\else\def\journalpre{}\fi\fi\fi
	\if\relax\detokenize{#3}\relax\def\insertjournal{\journalpre#2}\else\def\insertjournal{\journalpre\href{#3}{#2}}\fi
}
\newcommand{\doi}[1]{\if\relax\detokenize{#1}\relax\def\insertdoi{}\else\def\insertdoi{DOI: \href{http://dx.doi.org/#1}{#1}}\fi}
\newcommand{\arxiv}[2]{\if\relax\detokenize{#2}\relax\def\insertarxiv{}\else\def\insertarxiv{arXiv: \href{https://arxiv.org/abs/#1}{#1 [#2]}}\fi}
\newcounter{authors}
\newcounter{addresses}
\newcounter{keywords}
\newcommand{\addauthor}[2]{\csdef{author\arabic{authors}}{#1}\csdef{authoraddress\arabic{authors}}{#2}\stepcounter{authors}}
\newcommand{\addaddress}[1]{\csdef{address\arabic{addresses}}{#1}\stepcounter{addresses}}
\newcommand{\addkeyword}[1]{\csdef{keyword\arabic{keywords}}{#1}\stepcounter{keywords}}
\newcommand{\maketitlesub}{
	\newcounter{i}
	\newcounter{j}
	\noindent\textbf{%
		\Large\inserttitle\\[1em]
		\large\csuse{author0}$^{\csuse{authoraddress0}}$%
		\forloop{i}{1}{\value{i} < \value{authors}}{%
			, \csuse{author\arabic{i}}$^{\csuse{authoraddress\arabic{i}}}$%
		}
	}\\[1em]
	\normalsize
	\setcounter{j}{0}
	\forloop{i}{0}{\value{i} < \value{addresses}}{%
		\stepcounter{j}
		\ifnum\value{addresses}>1$^{\arabic{j}}$\fi\,\csuse{address\arabic{i}}
		\ifthenelse{\value{j}<\value{addresses}}{\\}{}
	}
	\ifx\insertemail\empty\\[1em]\else\\[0.5em]E-mail address: \insertemail\\[1em]\fi
	\textbf{Abstract:} \insertabstract
	\ifthenelse{\value{keywords}=0}{}{
		\\[1em]
		Keywords: \csuse{keyword0}%
			\forloop{i}{1}{\value{i} < \value{keywords}}{%
				; \csuse{keyword\arabic{i}}%
			}
	}
}
\renewcommand{\maketitle}{\iftwocolumn\oldtwocolumn[\maketitlesub\vspace{1.5em}]\else\maketitlesub\fi}

\usepackage[headsepline]{scrlayer-scrpage}
\clearpairofpagestyles
\ihead{%
	\ifx\insertarxiv\empty%
			\ifx\insertjournal\empty\else\textnormal\insertjournal\fi%
	\else%
		\ifx\insertdoi\empty%
			\ifx\insertjournal\empty\else\textnormal\insertjournal\fi%
		\else%
			\ifx\insertjournal\empty\else\textnormal\insertjournal\fi%
			\ifx\textnormal\empty\else\linebreak\textnormal\insertdoi\fi%
		\fi%
	\fi%
}
\ohead{%
	\ifx\insertarxiv\empty%
		\ifx\textnormal\empty\else\textnormal\insertdoi\fi%
	\else%
		\ifx\insertdoi\empty%
			\ifx\insertarxiv\empty\else\textnormal\insertarxiv\fi%
		\else%
			\ifx\insertarxiv\empty\else\linebreak\textnormal\insertarxiv\fi%
		\fi%
	\fi%
}
\ifoot{}
\ofoot{}

\usepackage[hang]{footmisc}
\let\oldthebibliography\thebibliography

\renewcommand\thebibliography[1]{
	\small
	\oldthebibliography{#1}
	\setlength{\parskip}{0pt}
	\setlength{\itemsep}{0pt plus 0.3ex}
}
\newcommand{\bstindent}{99}
\newcommand{\bstaddress}{}
\newcommand{\bstauthor}{}

\newcommand{\bstjournal}{}

\newcommand{\bstpublisher}{}
\newcommand{\bstinstitution}{}

\newcommand{\bsttitle}{}
\newcommand{\bstvolume}{}
\newcommand{\bstyear}{}
\newcommand{\bbland}{and}

\newcommand{\sref}[1]{section~\ref{#1}}

\newcommand{\fref}[1]{figure~\ref{#1}}
\newcommand{\Fref}[1]{Figure~\ref{#1}}
\newcommand{\tref}[1]{table~\ref{#1}}

\newcommand{\eref}[1]{(\ref{#1})}

\newcommand{\eqalign}[1]{\begin{aligned}#1\end{aligned}}